\documentclass[dvips]{article}
\usepackage{spie}
\usepackage{amsmath}
\usepackage{graphicx}

\input{psfig.sty}

\title{Development prospects and stability limits of the mid-IR Kerr-lens mode-locked lasers}

\author{V.L. Kalashnikov, E. Sorokin, and I.T. Sorokina
\skiplinehalf
Institut f\"ur Photonik, TU Wien, Gusshausstr. 27/387, A-1040 Vienna, Austria}

\authorinfo{Further author information: (Send correspondence to Dr. V.L. Kalashnikov)\\V.L. Kalashnikov:
E-mail: v.kalashnikov@tuwien.ac.at}

\begin{document}
  \maketitle

\begin{abstract}
The Kerr-lens mode locking ability and the ultrashort pulse
characteristics are analyzed numerically for the Cr-doped ZnTe,
ZnSe, ZnS active media. The advantages of these materials for the
femtosecond lasing within 2 - 3  $\mu$m spectral range are
demonstrated.
\end{abstract}


\section{INTRODUCTION}
\label{sect:intro}

Compact diode-pumped sources of the femtosecond pulses tunable
within the wavelength range between 2 and 3 $\mu$m are of interest
for various applications, such as laser surgery, remote sensing
and monitoring, spectroscopy of semiconductors etc. To date only
cryogenically operated Pb-salt diode lasers, optical parametrical
oscillators and difference-frequency convertors were available for
the operation in this spectral range. Therefore the possibility of
the direct mid-IR lasing from the new class of the
transition-metal doped chalcogenides
~\cite{DeLoach,Page,Hommerich,McKay} has attracted much attention.
The impressive advantages of these media are room-temperature
operation between 2 and 3 $\mu$m, possibility of direct diode
pumping, high emission and absorption cross-sections, negligibly
low excited-state absorption and, as consequence, low thermal load
(the basic laser material characteristics will be described in the
next section). The most remarkable examples of such lasers are
Cr$^{2+}$-doped ZnSe, ZnS and ZnTe. To date the following
achievements for these media have been demonstrated: 1) for
Cr:ZnSe CW operation with over 1.7 W power ~\cite{Wagner}, over
1100 nm tunability ~\cite{Sorokin1}, over 350 nm tunable
diode-pumped CW operation ~\cite{Sorokin2}, active mode locking
~\cite{Carrig} and active modulator assisted passive mode locking
~\cite{Sorokin3} were achieved; 2) for Cr:ZnS pulsed
~\cite{Page,Graham} and tunable CW operation ~\cite{Sorokin4} were
obtained. Cr:ZnTe, which is a member of the considered media
class, remains unexplored.

In spite of the numerous advantages, there exist some obstacles
for femtosecond pulse generation from these lasers. As they are
the semiconductors, i.e. possess a comparatively narrow band-gap,
the nonlinear refraction in the active crystal is large (see
below). Hence the self-focusing has a comparatively low threshold
which in the combination with the self-phase modulation produce a
tendency to the multiple pulse operation in the Kerr-lens mode
locking (KLM) regime ~\cite{Sorokin3,Kalash1}. Thus, there is a
need in the study of the KLM stability limits and methods of the
stability enhancement in the considered lasers. Moreover, the
large absorption and emission cross sections tend to the
stabilization of the CW operation, which prevents the KLM
self-start.

In this paper we present an analysis of the nonlinear refraction
in the Zinc-chalcogenides (by the example of ZnSe). In the
combination with the lasing properties this defines the main
requirements to the KLM optimization. Then, the results of the
numerical optimization of the KLM aimed at the multipulsing
suppression and taking into account a strong saturation of the
Kerr-lens induced fast absorber are presented. We demonstrate the
possibility of the few optical cycle pulse generation from the
Cr-doped Zinc-chalcogenodes. The problem of the self-starting KLM
is analyzed on the basis of the generalized momentum method taking
into account real-world laser configurations, gain guiding,
soft-aperture, thermo-lensing and other properties of the
considered laser materials. The presented models are quite general
and can be applied to the overall optimization of the different
KLM lasers.

  \section{DISTRIBUTED MODEL AND BASIC PARAMETERS}
\label{sect:model}

Simulation of the KLM can be based on the two quite different
approaches. First one supposes the full-dimensional modelling
taking into account the details of the field propagation in the
laser cavity ~\cite{Kalosha}. The minimal dimension of such models
is 2+1 and they allow the description of the spatio-temporal
dynamics of the ultrashort pulses and their mode pattern. Its main
disadvantages are a large number of the parameters resulting in
ambiguity of the optimization procedure and complexity of the
interpretation of the obtained results. Second approach is based
on 1+1 dimensional model in the framework of the so-called
nonlinear Ginzburg-Landau equation ~\cite{Kramer}, which describes
the KLM as an action of the fast saturable absorber governed by
the few physically meaningful parameters, viz., its modulation
depth $\gamma$ and the inverse saturation intensity $\sigma$. This
method allows the analytical realization in the week-nonlinear
limit ~\cite{Haus}, however in the general case the numerical
simulations are necessary. We shall base our analysis on the both
approaches.

In the beginning let us consider the master equation describing
the ultrashort pulse generation in the KLM solid-state laser:

\begin{equation} \label{LG}
\frac{{\partial a(z,t)}} {{\partial z}} = \left[ {\alpha  - \rho +
\left( {t_f^2 \frac{{\partial ^2 }}{\partial t^2 }  -
i\sum\limits_{m = 2}^N {\frac{{(-i)^m \beta _m }}
{{m!}}\frac{{\partial ^m }} {{\partial t^m }}} } \right) -
\frac{\gamma } {{1 + \sigma \left| {a(z,t)} \right|^2 }} - i\delta
\left( {\left| {a(z,t)} \right|^2  - \frac{i} {{\omega _0
}}\frac{\partial } {{\partial t}}\left| {a(z,t)} \right|^2 }
\right)} \right]a(z,t),
\end{equation}

\noindent where $a(z,t)$ is the field amplitude (so that $|a|^2$
has a dimension of the intensity), $z$ is the longitudinal
coordinate normalized to the cavity length (thus, as a matter of
fact, this is the cavity round-trip number), $t$ is the local
time, $\alpha$ is the saturated gain coefficient, $\rho$ is the
linear net-loss coefficient taking into account the intracavity
and output losses, $t_f$ is the group delay caused by the spectral
filtering within the cavity, $\beta_m$ are the $m$-order
group-delay dispersion (GDD) coefficients, $\delta$ = $l_g n_2
\omega _0 /c$ = $2\pi n_2 l_g/(\lambda _0 n)$ is the self-phase
modulation (SPM) coefficient, $\omega_0$ and $\lambda_0$ are the
frequency and wavelength corresponding to the gain band maximum,
$n$ and $n_2$ are the linear and nonlinear refraction indexes,
respectively, $l_g$ is the double length of the gain medium (we
suppose that the gain medium gives a main contribution to the
SPM). The last term in Eq.~\eqref{LG} describes the
self-steepening effect and for the simplification will be not
taken into account in the simulations. As an additional
simplification we neglect the stimulated Raman scattering in the
active medium ~\cite{Kalash2}. These two factors will be
considered hereafter.

The gain coefficient obeys the following equation:

\begin{equation} \label{gain}
\frac{{\partial \alpha (z,t)}} {{\partial t}} = \sigma _a \left(
{\alpha _{\max }  - \alpha (z,t)} \right)\frac{{I_p }} {{\hbar
\omega _p }} - \sigma _g \alpha (z,t)\frac{{\left| {a(z,t)}
\right|^2 }} {{\hbar \omega _0 }} - \frac{{\alpha (z,t)}} {{T_r
}}.
\end{equation}

\noindent Here $\sigma_a$ and $\sigma_g$ are the absorption and
emission cross-sections of the active medium, respectively, $T_r$
is the gain relaxation time, $I_p$ is the absorbed pump intensity,
$\omega_p$ is the pump frequency, $\alpha_{max}$ = $\sigma_g N_g
l_g$ is the maximum gain coefficient, $N_g$ is the concentration
of the active centers. The assumption $\tau_p \ll T_{cav}$
($\tau_p$ is the pulse duration, $T_{cav}$ is the cavity period)
allows the integration of Eq.~\eqref{gain}. Then for the
steady-state gain coefficient we have:

\begin{equation} \label{alpha}
\alpha  = \frac{{\alpha _{\max } \sigma _a I_p T_{cav} }} {{\hbar
\omega _p \left( {\frac{{\sigma _a I_p T_{cav} }} {{\hbar \omega
_p }} + \frac{E} {{E_s }} + \frac{{T_{cav} }} {{T_r }}} \right)}},
\end{equation}

\noindent where $E_s$ = $\hbar \omega _p /\sigma _g$ is the gain
saturation energy flux, $E$ = $ \int_{ - T_{cav} /2}^{T_{cav} /2}
{\left| {a(t)} \right|^2 } dt $ is the pulse energy.

For the numerical simulations in the framework of the distributed
model it is convenient to normalize the time and the intensity to
$t_f$ = $\lambda_0 ^2 /(\Delta \lambda c) $ and $1/\delta$,
respectively ($\Delta \lambda$ is the gain bandwidth). The
simulation were performed on the $2^{12} \times 10^4$ mesh. Only
steady-state pulses were considered. As the criterion of the
steady-state operation we chose the peak intensity change less
than 1\% over last 1000 cavity transits.

The KLM in the considered model (\ref{LG}, \ref{alpha}) is
governed by the only four basic parameters: $\alpha - \rho$,
$\beta_2$, $\gamma$, and $\sigma$. This allows unambiguous
multiparametric optimization. In the presence of the higher-order
dispersions, the additional $\beta_m$ parameters appear. This
complicates the optimization procedure, but keeps its physical
clarity.

Now let us consider the basic active media parameters.

\begin{table}[hbt]
\caption{\label{params-exp} Material parameters of the Cr-doped
Zinc-chalcogenides. }

\begin{center}
\begin{tabular}{|c|c|c|c|c|c|c|c|c|}
\hline
 Medium  & $\lambda_0$, $\mu$m & $\Delta \lambda$, nm & $\lambda_a$, $\mu$m & $\sigma_a$, 10$^{-19}$ cm$^2$ & $\sigma_g$, 10$^{-19}$ cm$^2$ & $n$  & $n_2$, 10$^{-13}$ esu & $T_r$, $\mu$s \\
\hline
 Cr:ZnSe &         2.5         &         880          &        1.61         &              8.7              &               9               & 2.44 &          29 -- 92          &      6-8      \\
\hline
 Cr:ZnS  &        2.35         &         800          &        1.61         &              5.2              &              7.5              & 2.3  &          16 -- 30           &     4-11      \\
\hline
 Cr:ZnTe &         2.6         &         800          &        1.61         &              12               &              20               & 2.71 &          80 -- 150          &       3       \\
\hline
\end{tabular}
\end{center}
\end{table}

\begin{figure}
   \begin{center}
   \begin{tabular}{c}
   \psfig{figure=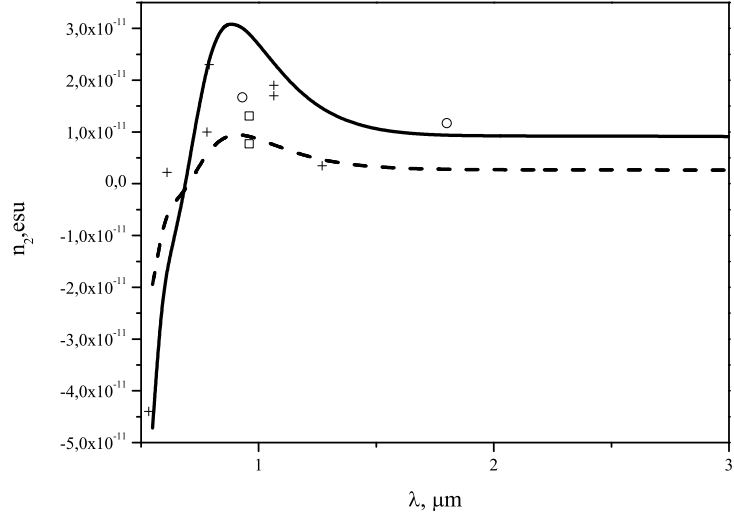,height=8cm}
   \end{tabular}
   \end{center}
   \caption[Fig. 1]
   { \label{disp}
Calculated wavelength dependence of $n_2$ for ZnSe. Solid and
dashed curves present the maximum and minimum estimations based on
Eq. (\ref{n2}). Points give the experimental data for
polycrystalline samples, squares -- for monocrystals with
different orientations, crosses -- unknown crystalline
structures.}
   \end{figure}

It should be noted that the $n_2$ values were experimentally
measured at the wavelengthes which are smaller than $\lambda_0$
(see Fig. ~\ref{disp}). Therefore we have to estimate theirs at
the generation wavelength. Such estimation can be obtained from
the two ~\cite{Bahae} or four ~\cite{Hutchings} bands  models of
semiconductor. The most simple estimation results from the former
model, which gives the following formula:

\begin{equation} \label{n2}
n_2 (esu) = K\frac{{G(\hbar \omega /E_g )\sqrt {E_p } }} {{n E_g^4
}},
\end{equation}

\noindent where $K$ = (0.5 - 1.5)$\times10^{-8}$ and $E_p$ = 21 eV
are the material independent constants, $E_g$ is the band-gap
width in eV, $G$ is the form-factor. Using for $K$ the values
0.86--1.5 $\times 10^{-8}$ and for $E_p$ the values 21--24 eV, we
obtained the estimations presented in Table~\ref{params-exp}. The
calculated $n_2$ dispersion for ZnSe is shown in Fig. ~\ref{disp}
~\cite{Kalash5}.

\begin{figure}
   \begin{center}
   \begin{tabular}{c}
   \includegraphics{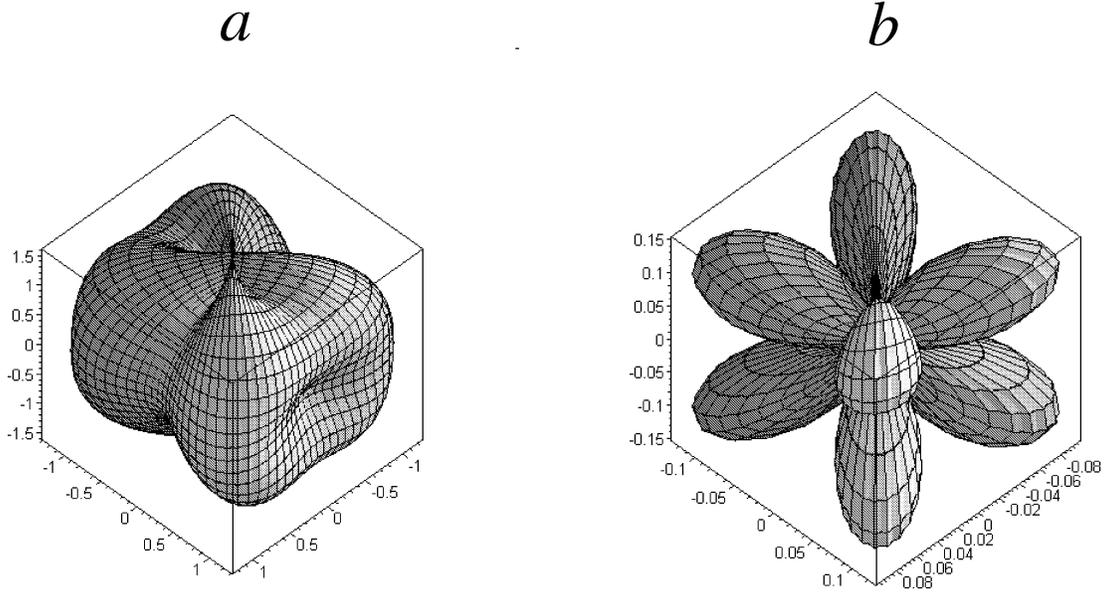}
   \end{tabular}
   \end{center}
   \caption[Fig. 2]
   { \label{orient}
(\textit{a}): Calculated orientational dependence of the
third-order $n_2$ for ZnSe ~\cite{Hutchings2} scaled to
$n_2^{[001]}$ and (\textit{b}): orientational dependence of the
second-order nonlinear coefficient scaled to its maximum value.}
   \end{figure}

We note that the semiconductor nature of these active media
results in the extremely high nonlinear refraction coefficients in
the comparison with Ti:sapphire (1.2$\times10^{-13}$ esu). Also,
we have to note that there exists the orientational dependence of
$n_2$, which can be especially large for the birefringent ZnS. But
even for the cubic ZnSe, the orientational variation of $n_2$ can
reach 70\% (see Fig.~\ref{orient}).

\begin{figure}
   \begin{center}
   \begin{tabular}{c}
   \psfig{figure=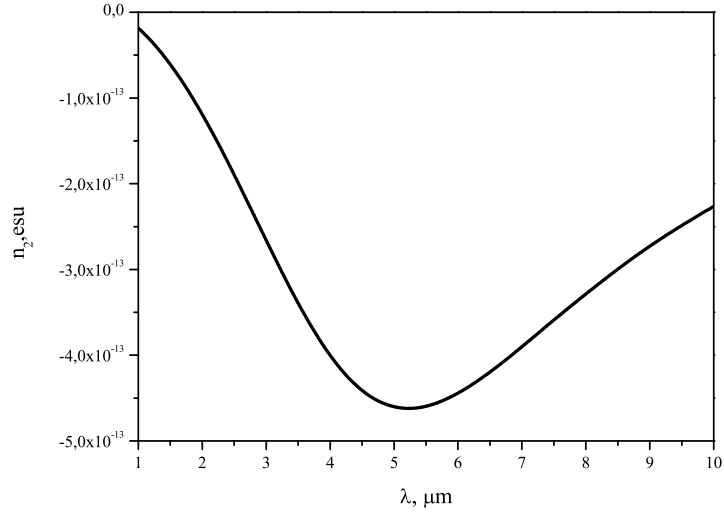,height=8cm}
   \end{tabular}
   \end{center}
   \caption[Fig. 3]
   { \label{cascad}
Calculated wavelength dependence of the cascaded $n_2$ for ZnSe.}
   \end{figure}

In addition to the large $n_2$ the considered media possess the
strong quadratic-nonlinearity due to absence of the center of
inversion (for ZnSe and ZnTe). For example, $d_{36}$ = 33 pm/V for
ZnSe. However, as a result of the cubic structure, there exists
the strong orientational dependence of the effective second-order
nonlinearity (see Fig.~\ref{orient}), so that the maximum
effective nonlinear coefficient is only 0.17$d_{36}$. The coherent
length for this media is only 23 $\mu$m that strongly reduces an
efficiency of the second-harmonic generation. Nevertheless, the
cascaded second-order process can contribute to the effective
$n_2$ ~\cite{Sutherland,schiek}. Our calculation demonstrate (see
Fig. ~\ref{cascad} ~\cite{Kalash5}) that this contribution is
small in the comparison with the third-order $n_2$.

As a result, the simulation parameters corresponding to the above
introduced normalizations are summarized in Table~\ref{th}. $ P =
\sigma _{a} T_{cav} I_p /\left( {\hbar \omega _0 } \right)$,
$\epsilon$ = $t_f E_s^{-1}/\delta$; $P_{cr}$=$2 n_0/(n_2 k^2)$ and
$I_{sat}$=$h \nu_0/(\sigma_g T_r)$ are the critical power of the
self-focusing (beyond the parabolical approximation for the
Gaussian beam) and the gain saturation intensity, respectively.

\begin{table}[hbt]
\caption{\label{th} Simulation parameters. $l_g$=2$\times$0.3 cm,
$T_{cav}$=10 ns, 2 W pump power, 100$\times$100 $\mu$m$^2$ pump
mode. We used the average value of $n_2$ and the minimum of
$T_r$.}
\begin{center}
\begin{tabular}{|c|c|c|c|c|c|c|c|c|}
\hline
 Medium  & $\alpha_{max}$ &  $n_2$, 10$^{-16}$ cm$^2$/W & $P_{cr}$, MW   & $\delta$, 10$^{-12}$ cm$^2$/W & $\epsilon$, 10$^{-4}$ & $I_{sat}$, kW/cm$^2$ & $t_f$, fs & P, 10$^{-3}$ \\
\hline
 Cr:ZnSe &       5        &       105             &     0.75     &              87               &          4.9      &    15     &    3.8   &     1.1      \\
\hline
 Cr:ZnS  &       5        &         42            &     1.5     &              32               &          10       &     28    &    3.7   &     0.62     \\
\hline
 Cr:ZnTe &       5        &       178             &     0.52     &               314              &          3.8      &   13      &    4.5  &      1.9      \\
\hline
\end{tabular}

\end{center}
\end{table}

  \section{DISTRIBUTED MODEL: RESULTS AND DISCUSSION}
\label{sect:results}

First of all we have to consider the meaning of the optimization
procedure. The numerical simulations demonstrate that there exist
some values of the saturation parameter $\sigma$, which provide
the near chirp-free pulse generation. These values of $\sigma$ can
be considered as ``optimal''. However, there is an additional
factor, which has to be taken into account: the pulse shortening
is possible by the $\sigma$ growth and $\beta_2\rightarrow$0. We
consider the values of $\beta_2$ and $\sigma$ as optimal if they
correspond to the generation of the shortest pulse. The main
obstacle on the way of the pulse shortening is the multiple pulse
generation~\cite{Kalash1}. The strong tendency to the multipulsing
results from 1) $E_s^{-1}$ decrease, 2) $P$ increase, 3) fast
absorber saturation favored by the large $\sigma$. The relatively
large values of $\sigma_g$ and $\lambda_0$ for the considered
media result in the increase of the first parameter. However, the
large absorption cross sections and SPM coefficients increase two
later parameters. If the growth of the first parameter corresponds
to the gain saturation and thereby to the stabilization of the
pulse against the laser continuum growth, the larger $P$ and
$\sigma$ initiate the rise of the continuum and the excitation of
the perturbations inside the ultrashort pulse~\cite{Kalash3}.
Hence, the optimization means the multiple pulses suppression
allowing the pulse shortening.

Additional limitations on the pulse shortening result from the
achievable values of the saturation parameter $\sigma$. In the KLM
lasers this parameters is governed by the cavity alignment: the
shift towards the cavity stability zone increases $\sigma$ (see
next section). So, the highest values of $\sigma$ are reached in
the immediate vicinity of the cavity stability boundary. This
requires a too thorough cavity optimization and can not be
considered as operational. Hence, the optimization aimed at the
pulse shortening is based on the variation of all four parameters
of master Eq.~\eqref{LG} and is constrained by the above described
reasons.

\begin{figure}
   \begin{center}
   \begin{tabular}{c}
   \includegraphics[width=12cm]{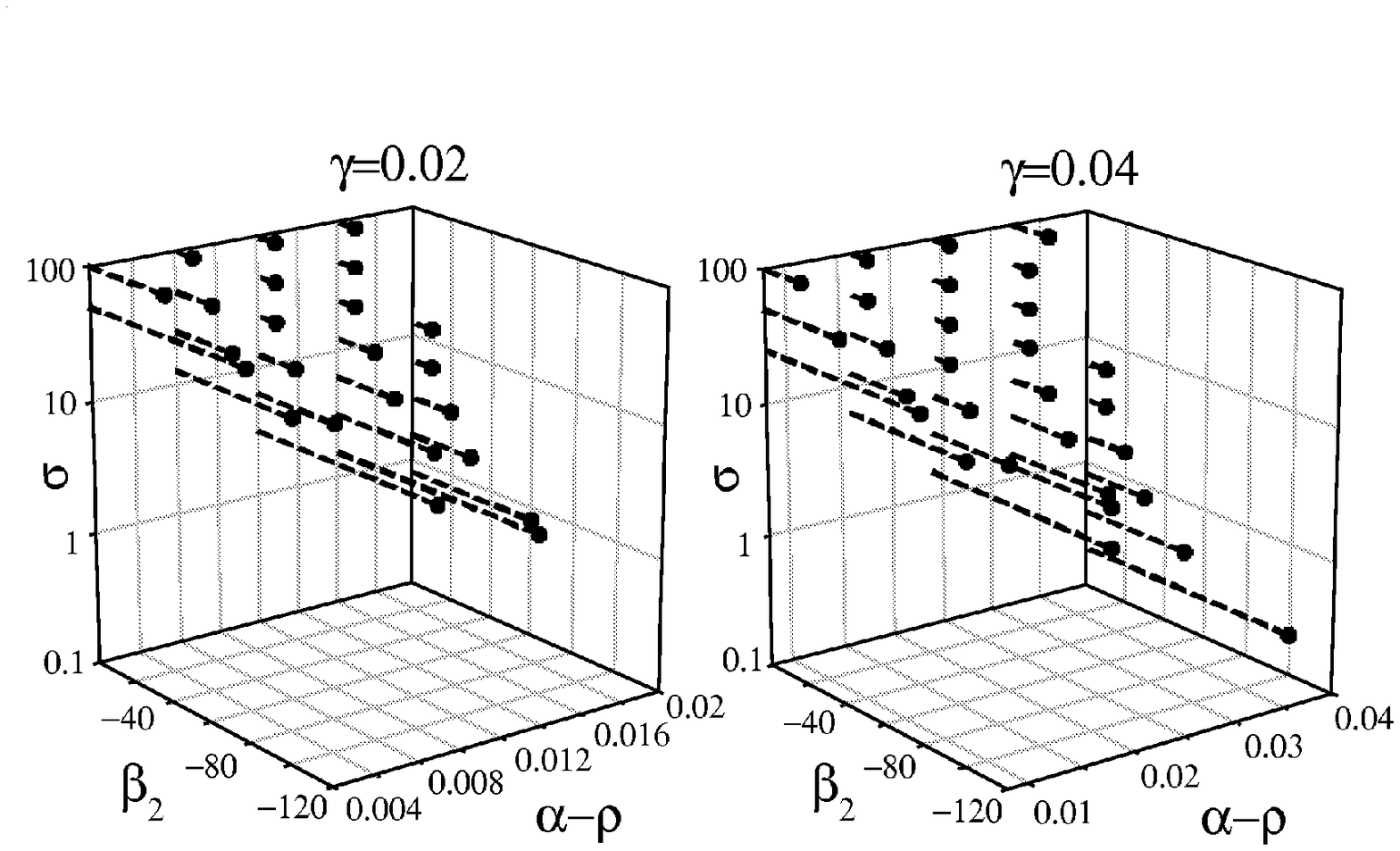}
   \end{tabular}
   \end{center}
   \caption[Fig. 4]
   { \label{3d-1}
Stability boundary for the single pulse operation. Dimensional GDD
[fs$^2$] = $\beta_2 \times t_f^2$}
   \end{figure}

Fig.~\ref{3d-1} shows the parameters of Eq.~\eqref{LG}, which for
the fixed $\alpha-\rho$, $\gamma$ and $\sigma$ correspond to the
minimum achievable $\beta_2$, where the pulse width has a minimum.
The further $\beta_2$ decrease results in the multipulse
operation. Thus, the points in Fig.~\ref{3d-1} lie on the boundary
of the stable single pulse operation. There is a set of the
general tendencies characterizing this boundary.

\begin{figure}
   \begin{center}
   \begin{tabular}{c}
   \psfig{figure=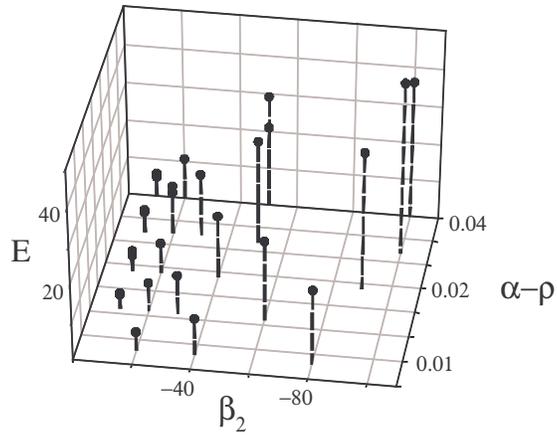,height=6cm}
   \end{tabular}
   \end{center}
   \caption[Fig. 5]
   { \label{En}
Pulse energy E on the stability boundary for $\gamma$=0.04.}
   \end{figure}

There exists a limited on $\alpha-\rho$ range of the stable single
pulse operation, which expands as a result of the $\sigma$ and
$\gamma$ growth. The increase of $\sigma$ shifts this region
towards the smaller $\alpha-\rho$. However, when $\sigma >$100 the
transition to multipulsing is possible. As every point in
Fig.~\ref{3d-1} corresponds to the fixed dimensionless $E$ (see
Fig.~\ref{En}), the choice of the appropriate level of
$\alpha-\rho$ for the fixed laser configuration (i.e. fixed
$\gamma$, $\sigma$ and $\beta_2$) is realized by the variation of
$P$ (pump, see Eq.~\eqref{alpha}) as well as $\rho$ (output loss).
It is possible also to change $\alpha$ by the change of
$\alpha_{max}$ due to variation of the crystal length or the
active ions concentration or by the change of $T_r$ due to
variation of the active ions concentration. Note also, that the
$l_g$ decrease increases $\epsilon$ (due to the $\delta$
decrease), which describes the ``strength'' of the gain saturation
relatively the SPM. Hence, $\epsilon$ increase expands the
stability region towards the higher pump and allows the higher
pulse energies (because they $\propto 1/\delta$).

The $E$ decrease for the fixed $\alpha-\rho$, which takes a place
for $\beta_2\rightarrow$0 (Fig.~\ref{En}), has to be accompanied
by $P$, $\alpha_{max}$, $T_r$ decrease or by $\epsilon$, $\rho$
increase in order to prevent from the continuum amplification. The
last is the main source of the pulse destabilization and
suppresses the single pulse operation in the vicinity of zero GDD.
Hence, the pulse generation for $\alpha-\rho-\gamma>$0 is not
possible. $\alpha-\rho-\gamma=$0 corresponds to the specific
hybrid regime with the coexistent pulse and continuum
~\cite{Akhmediev}.

Since the approach of the GDD to zero has to be accompanied by the
pump decrease, this results in the growth of the $\sigma$ required
for the pulse stabilization (Fig.~\ref{3d-1}). This can demand too
thorough cavity alignment. Moreover, for the large $\sigma$ we
need the larger minimum $|\beta_2|$ providing the single pulse
operation so that the dependence of the minimum $|\beta_2|$ on the
$\sigma$ for the fixed $\alpha-\rho$ has a parabolic-like form
~\cite{Kalash3}.

The most interesting features are the shift of the stability
region towards the smaller $\sigma$ (Fig.~\ref{3d-1}) and the
pulse shortening as a result of the $\gamma$ increase. For
example, the minimum pulse duration $\tau_p$ for $\gamma=$0.03 is
7$t_f$ whereas for $\gamma=$0.05 it is 5$t_f$ (this is 19 fs for
Cr:ZnSe and Cr:ZnS and 23 fs for Cr:ZnTe). The bad news here is
the need for the hard-aperture KLM to provide the larger
modulation depth. This reduces the KLM self-starting ability.

\begin{figure}
   \begin{center}
   \begin{tabular}{c}
   \psfig{figure=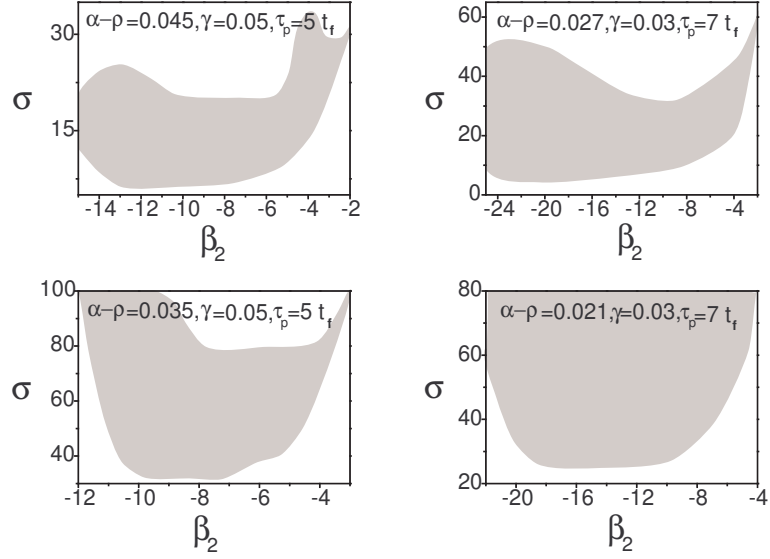,height=8cm}
   \end{tabular}
   \end{center}
   \caption[Fig. 6]
   { \label{zone}
Regions of the $\sigma$ and $\beta_2$ parameters providing the
shortest pulses.}
   \end{figure}

The regions of the parameters allowing the shortest pulses are
shown in Fig.~\ref{zone}. The $\alpha-\rho$ increase reduces the
minimum $\sigma$ parameter producing the shortest pulses. However,
the region of their existence shortens on $\sigma$.

   \begin{figure}[h]
   \begin{center}
   \begin{tabular}{c}
   \psfig{figure=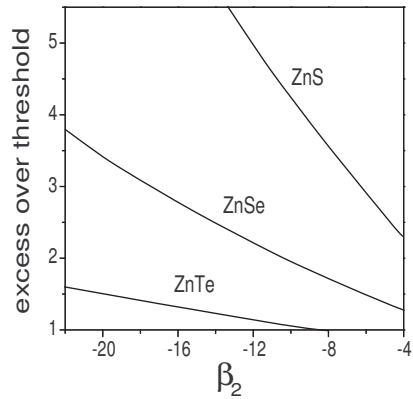,height=6cm}
   \end{tabular}
   \end{center}
   \caption[Fig. 7]
   { \label{thresh}
Pump threshold of KLM related to that for CW. Pulse duration is
$\tau_p=$7 $t_f$. $\alpha=$0.071, $\rho=$0.05, $\gamma=$0.03,
$\alpha_{max}=$5, other parameters correspond to Table~\ref{th}.}
   \end{figure}

Let us consider the concrete parameters of Table~\ref{th}. The
pump thresholds allowing 7$t_f$ pulse durations are shown in
Fig.~\ref{thresh}. The threshold decreases from Cr:ZnS through
Cr:ZnSe to Cr:ZnTe that results from the $\epsilon$ decrease. This
means that the SPM becomes stronger relatively the gain saturation
under this transition. As a result, the tendency to the pulse
destabilization intensifies and this demands to reduce the
intracavity power by means of the pump decrease.

   \begin{figure}[h]
   \begin{center}
   \begin{tabular}{c}
   \psfig{figure=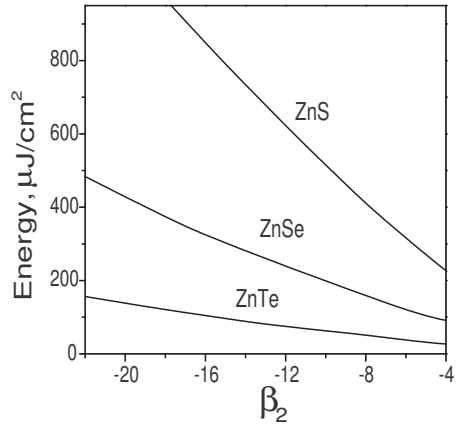,height=6cm}
   \end{tabular}
   \end{center}
   \caption[Fig. 8]
   { \label{energy}
Intracavity pulse energy fluxes. The parameters correspond to
Fig.~\ref{thresh}.}
   \end{figure}

Thus, the $\epsilon$ decrease reducing KLM threshold turns in the
intracavity pulse energy decrease (Fig.~\ref{energy}). The highest
value of $\epsilon$ for Cr:ZnS resulting from the large $\sigma_g$
in the combination with comparatively small $\delta$ produces the
stabilization of the shortest pulses with the highest energies.

Note, that the larger absorption cross-section for Cr:ZnTe results
in the highest absorbed pump energy for the fixed pump intensity
and mode area. This is a positive factor for the KLM threshold
lowering. However, this can be a negative factor, when the SPM is
the source of the pulse destabilization because the additional
efforts for the intracavity power decrease are necessary (see also
next section).

At last, we consider the contribution of the third-order
dispersion, which can be large for the lasers under consideration.
There are the technological troubles in the use of the
chirped-mirror technique for the dispersion compensation in the
mid-IR due to high value of $\lambda_0$. Therefore the usual
technique utilizing the prisms for the dispersion control can be
useful in the considered situation. As a result, the third-order
net-dispersion coefficient $|\beta_3|$ increases.

\begin{figure}[h]
   \begin{center}
   \begin{tabular}{c}
   \includegraphics[width=12cm]{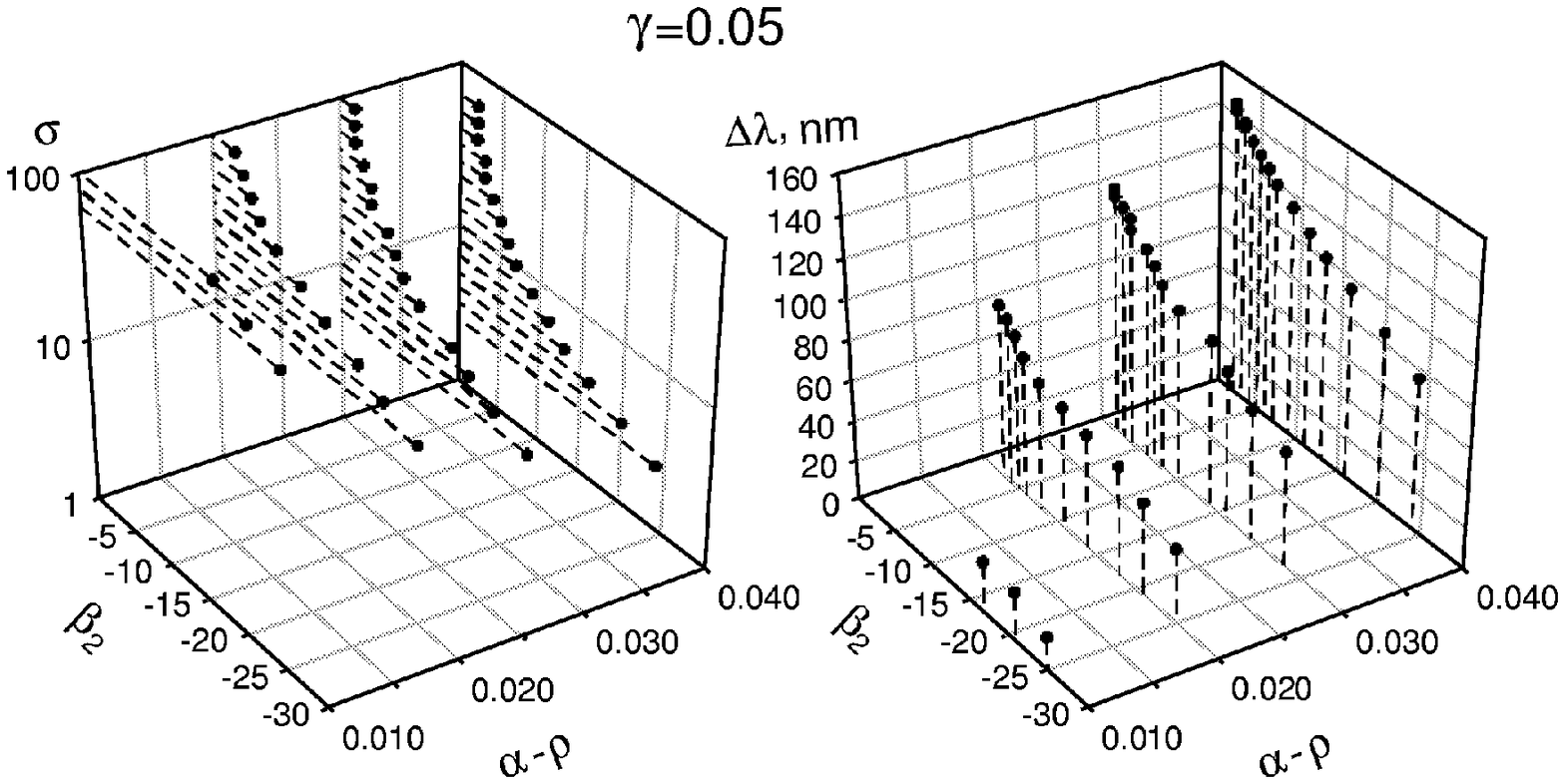}
   \end{tabular}
   \end{center}
   \caption[Fig. 9]
   { \label{tod}
Stability boundary for the single pulse operation in the presence
of the third-order dispersion and the Stokes spectral shift of the
pulse spectrum on this boundary.}
   \end{figure}

For the simulation we choose $\beta_3=$-5900 fs$^3$. As it can be
seen from Fig.~\ref{tod}, the shape of the stability boundary does
not change in the comparison to $\beta_3=$0 ~\cite{Kalash4}.
However, the minimum pulse duration increases from 5$t_f$ for
$\beta_3=$0 to 9$t_f$ (34 fs for Cr:ZnS, Cr:ZnSe and 40 fs for
Cr:ZnTe). The additional effect is the pronounced (up to 140 nm)
Stokes shift of the peak wavelength (Fig.~\ref{tod} shows this
shift on the stability boundary for Cr:ZnSe and Cr:ZnS). This
shift is typical also for such IR lasers as Cr:LiSGaF, Cr:LiSAF,
Cr$^{4+}$:YAG ~\cite{Kalash2,Kalash4} and can reduce the pulse
energy due worse overlap between gain band and pulse spectrum.
However, for the media under consideration the wavelength shift is
small in comparison with the full gain band width and the energy
decrease is not critical.

\section{TIME-SPATIAL MODEL: SELF-START ABILITY OF THE KERR-LENS MODE LOCKING}

In the previous section we considered the ultrashort pulse
stability on the basis of the distributed 1+1-dimensional model.
However, this model can not provide answers to the basic
questions: what is the self-start ability of the Kerr-lens mode
locking and what has the real-world laser configuration to be?

These questions require an analysis on the basis of the
time-spatial model. The simplest way is the assumption of the
Gaussian spatial distribution for the laser beam reducing problem
to the 1+2-dimensions. The free-space propagation of the Gaussian
beam can be considered on the basis of the usual ABCD-matrix
formalism, while the propagation inside the nonlinear active
medium is described by the following equation:

\begin{gather} \label{self}
  \frac{{\partial a\left( {z,r,t} \right)}}
{{\partial z}} - i\left[ {\frac{{2\vartheta r^2 a\left( {z,r,t}
\right)}} {{w_p^2 }} - \frac{{\frac{{\partial a\left( {z,r,t}
\right)}} {{\partial r}} + \frac{{r\partial ^2 a\left( {z,r,t}
\right)}} {{\partial r^2 }}}} {{2kr}} + \beta_2 '^2
\frac{{\partial ^2 a\left( {z,r,t} \right)}}
{{\partial t^2 }}} \right]a\left( {z,r,t} \right) + i\chi \left| {a\left( {z,r,t} \right)} \right|^2 a\left( {z,r,t} \right) =  \\
 \alpha \exp \left( { - \frac{{2r^2 }}
{{w_p^2 }}} \right)a\left( {z,r,t} \right) + t_f '^2
\frac{{\partial ^2 a\left( {z,r,t} \right)}} {{\partial t^2 }}.
\nonumber
\end{gather}

\noindent Here $\beta_2 '$ and $t_f '$ are the group-velocity
dispersion and the inverse group-velocity delay coefficients (for
ZnSe we used $\beta_2 '^2$=2054 fs$^2$/cm and $t_f '$=13 fs/cm).

The left-hand side of Eq. (\ref{self}) describes the
non-dissipative factors: thermo-lensing \cite{schroeder}
($\vartheta$=$k\frac{{dn_0 }} {{dT}}\zeta P_a\exp \left( { - \zeta
z} \right)$/ $\left( {4\pi n_0 \kappa _{th} } \right)$, $k$ is the
wave number, $\frac{{dn_0 }} {{dT}}$ is the coefficient of the
refractive index thermo-dependence (5.35$\times 10^{-5} K^{-1}$
for ZnSe), $\zeta$ is the loss coefficient at the pump wavelength,
$P_a$ is the pump power, $\kappa _{th}$ is the thermo-conductivity
coefficient (0.172 $W K^{-1} cm^{-1}$ for ZnSe)); diffraction (in
the cylindrically symmetrical case); group-velocity dispersion and
self-phase modulation (providing self-focusing for radially
varying beam, $\chi$=$n_2 k/n_0$). The right-hand side of Eq.
(\ref{self}) describes the dissipative factors inside the gain
medium: radially varying gain (providing gain guiding and soft
aperture action, $\alpha$ and $w_p$ are the saturated gain
coefficient and the pump beam size, respectively); spectral
filtering caused by the gain band profile.

It is convenient to rewrite Eq. (\ref{gain}) in the following way:

\begin{equation} \label{gain2}
\alpha  = \frac{{2 \alpha _{\max } \sigma _a P_a T_r }}{{\hbar
\omega _p \pi w_p^2 \left( {\frac{{2 \sigma _a P_a T_r }}{{\hbar
\omega _p \pi w_p^2}} + \frac{{2\upsilon P_g}}{{\pi w^2 I_s
}}\frac{{\tau _p }}{{T_{cav} }} + 1} \right)}},
\end{equation}

\noindent where $\upsilon$=$E \pi w^2/(2 P_g \tau_p)$ ($P_g$ is
the generation power), $w$ is the generation mode beam size.
$\upsilon$=$\sqrt {{\raise0.7ex\hbox{$\pi $} \!\mathord{\left/
 {\vphantom {\pi  2}}\right.\kern-\nulldelimiterspace}
\!\lower0.7ex\hbox{$2$}}}$ for the pulse with the Gaussian
time-profile, 2 for the $sech$-shaped pulse and 1 for the CW (in
the latter case $\tau_p$=$T_{cav}$).

The crucial simplification in the analysis of Eq. (\ref{self}) is
based on the so-called aberrationless approximation: the
propagating field has the invariable spatial-time profile, which
is described by the set of the $z$-dependent parameters. In the
non-dissipative case this approximation allows the variational
approach providing rigorous description of the Gaussian beam
propagation outside the parabolical approximation
~\cite{Anderson1,Anderson2,Anderson3}.

In the dissipative case we have to use the momentum method
~\cite{Vlasov}. However, in contrast to Refs.
~\cite{Herrmann1,Herrmann2} we shall consider the momentums
resulting from the symmetries of Eq. (\ref{self}). The $a \to
a\exp \left( {i\phi } \right)$ invariance, the transverse and time
translating invariance suggest the following momentums
~\cite{Akhmediev2}:

\begin{gather} \label{momentum} \nonumber
  T_{m,n}  = \iint\limits_\infty  {r^m t^n \left| a \right|^2 drdt,} \hfill \\
  J_{m.n}  = \iint\limits_\infty  {r^m t^n \left( {a\frac{{\partial a^* }}
{{\partial t}} - a^* \frac{{\partial a}}
{{\partial t}}} \right)drdt,} \hfill \\
  M_{m.n}  = \iint\limits_\infty  {r^m t^n \left( {a\frac{{\partial a^* }}
{{\partial r}} - a^* \frac{{\partial a}} {{\partial r}}}
\right)drdt.} \nonumber
\end{gather}

Like the variational approach we can substitute to Eqs.
(\ref{momentum}) the trial solution describing the ultrashort
pulse. If we take the Gaussian time-spatial profile $a\left(
{z,r,t} \right) = W(r)\exp \left( {G(r)} { - \frac{{r^2 }}
{{2w(r)'^2 }} + i b(r) r^2  - \frac{{t^2 }} {{\tau(r) ^2 }} +
i\psi(r) t^2 } \right)$ ($W(r)$ is the complex amplitude, $2 w
'^2$=$w^2$, $G(r)$ is the pulse amplification parameter excepting
the geometrical magnification for the Gaussian beam), the
equations describing the evolution of the pulse and beam
parameters are \cite{Kalash6}:

\begin{gather} \nonumber
\frac{{dw'}}{{dz}} =  - \frac{2}{k}w'\left( z \right)b\left( z
\right) - \frac{{2\alpha w'\left( z \right)^3 }}{{w_p^2 \left( {1
+ \frac{{2w'\left( z \right)^2 }}{{w_p^2 }}}
\right)^{{\raise0.7ex\hbox{$3$} \!\mathord{\left/
 {\vphantom {3 2}}\right.\kern-\nulldelimiterspace}
\!\lower0.7ex\hbox{$2$}}} }},\\ \nonumber \frac{{d\tau \left( z
\right)}}{{dz}} = \left[ {\frac{2}{{\tau \left( z \right)}} -
2\tau \left( z \right)^3 \psi \left( z \right)^2 } \right] t_f^2 +
2 \beta_2 '^2 \tau \left( z \right)\psi \left( z \right),\\
\label{iter} \frac{{dG\left( z \right)}}{{dz}} = \frac{\alpha }{{1
+ \frac{{2w'\left( z \right)^2 }}{{w_p^2 }}}} - \frac{{2t_f^2
}}{{\tau \left( z \right)^2 }} - \beta_2 '^2 \psi \left( z
\right),\\ \nonumber
 \frac{{db\left( z \right)}}{{dz}} =
\frac{{2\vartheta }}{{w_p^2 }} + \frac{{2b\left( z \right)^2 }}{k}
+ \frac{{\sqrt 2 P_0 e^{2G\left( z \right)} }}{{\pi P_{cr} k
w\left(
z \right) '^4 }} - \frac{1}{{2kw'\left( z \right)^4 }},\\
\nonumber
 \frac{{d\psi \left( z \right)}}{{dz}} =
2 \beta_2 '^2 \left( {\frac{1}{{\tau \left( z \right)^4 }} - \psi
\left( z \right)^2 } \right) - \frac{{8t_f^2 \psi \left( z
\right)}}{{\tau \left( z \right)^2 }} + \frac{{2 P_0 e^{2G\left( z
\right)} }}{{\pi P_{cr} w\left( z \right) '^2 \tau \left( z \right)^2 }},\\
\nonumber P_g  = P_0 e^{2G\left( z \right)},
\end{gather}

\noindent where $P_0$ and $w_0 '$ are the power and the beam size
before the active medium, respectively.

Eq. (\ref{iter}) in the combination with the ABCD-formalism allows
defining the stability regions for CW, single- and multiple
pulsing (the latter requires the trivial generalization of the
trial function). Such regions can predict the real-world laser
configurations providing the self-starting Kerr-lens mode locking.

\begin{figure}[h]
   \begin{center}
   \begin{tabular}{c}
   \psfig{figure=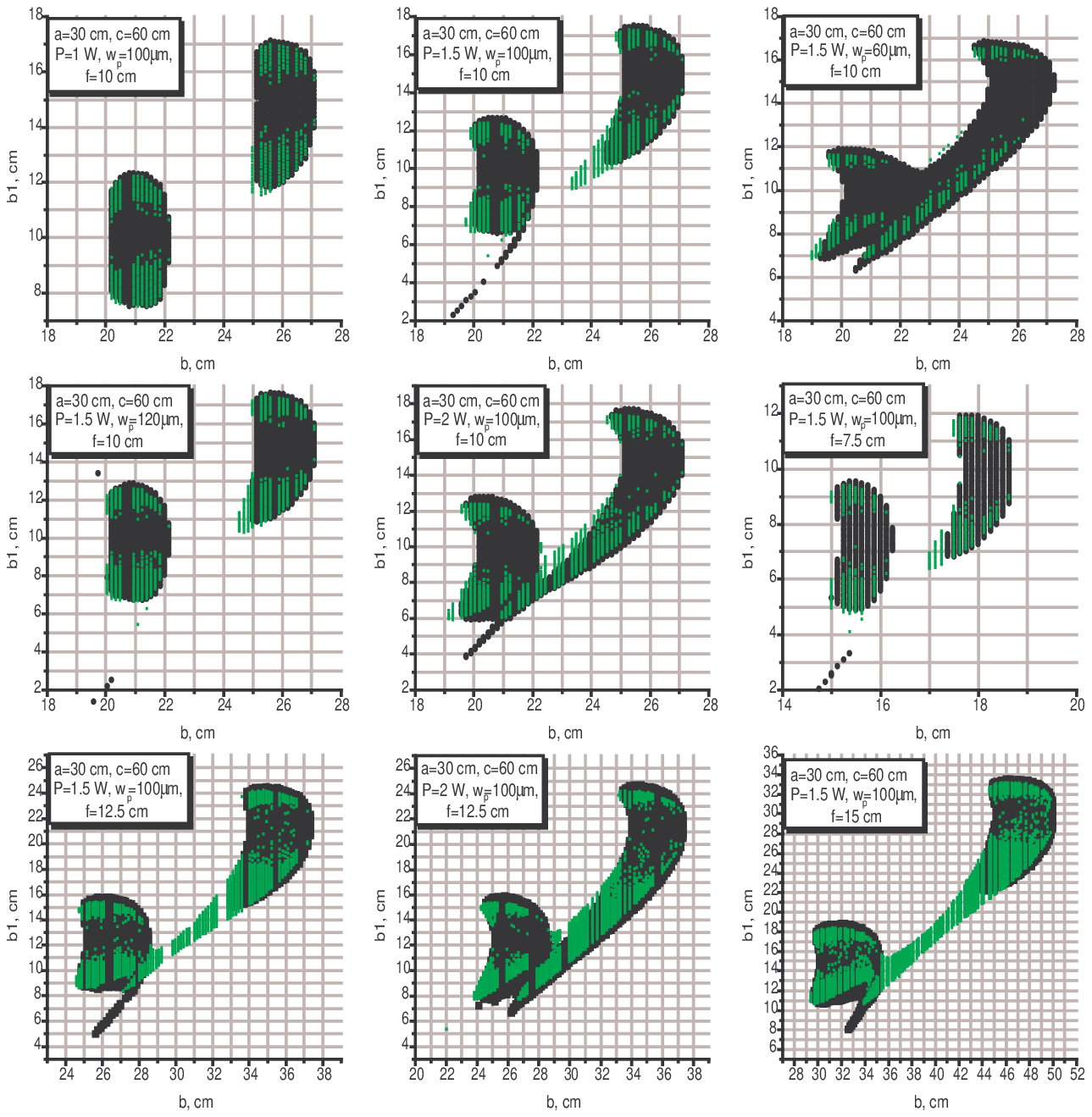,height=18cm}
   \end{tabular}
   \end{center}
   \caption[Fig. 10]
   { \label{zones}
Stability ranges for the CW (black filled circles) and mode
locking (open green circles). $a$ and $c$ are the cavity arms, $b$
is the folding distance, $b1$ is the distance of the active medium
facet from the curved mirror forming the shorter cavity arm, $f$
is the focal length of the curved mirrors. $\rho$=0.05.}
   \end{figure}

As it is seen from Table 2 the main features of the considered
active media are the comparatively low $P_{cr}$ (e.g. $P_{cr}$=2.6
MW for Ti:sapphire) and the extremely low $I_{sat}$ (300 kW for
Ti:sapphire). This causes the stabilization of the CW against the
mode locking due to the strong gain saturation. However, as it is
seen from Eqs. (\ref{gain2}, \ref{iter}) (note that the gain
saturation is intensity-dependent, but the self-focusing is
power-dependent), we can enhance the tendency to the mode locking
due to the suppression of the gain saturation. This can be
achieved by the growth of the generation mode within the active
medium owing to, for example, the cavity shortening or the
decrease of the folding mirrors curvature (Fig. ~\ref{zones}). At
the same time we have to keep the moderate level of the absorbed
pump power and to use the asymmetrical cavity in order to separate
the stability zones. The preferable configurations correspond to
the pulse operation without the CW. This occurs between the CW
stability zones (see Fig. ~\ref{zones}).

  \section{CONCLUSION}
\label{sect:conclusion}

In conclusion, we presented the models, which can be usable for
the numerical optimization of the KLM lasers. On the basis of
these models, the KLM abilities of the Cr-doped Zinc-chalcogenides
were estimated. It was shown, that the strong SPM inherent to
these media and destabilizing the single pulse operation can be
overcomed by the choice of the appropriate GDD, pump, modulation
depth and saturation parameter of the Kerr-lensing induced fast
saturable absorber. As a result, the Cr:ZnTe possesses the lowest
KLM threshold, however strong SPM constrains the achievable pulse
power for this laser. The best stability for the highest energy
flux and the shortest pulse duration (19 fs) are achievable in
Cr:ZnS. Cr:ZnSe. The presence of the third-order dispersion
increases the minimum achievable pulse durations up to 34 - 40 fs
and causes the strong (up to 140 nm) Stokes shift of the
generation wavelength. However, the latter effect does not reduce
noticeably the pulse energy. The Kerr-lens mode locking self-start
ability of the considered active media is reduced by the strong
gain saturation so that the cavity tuning providing the ultrashort
pulsing differs essentially from that for the Ti:sapphire. We have
to avoid over-pumping and over-shortening of the generation mode.
As a result, the short and asymmetrical cavities with the
comparatively large curvature of the folding mirrors are
preferable. On the whole, the Cr-doped Zinc-chalcogenides have the
prospects for the sub-50 fs generation that amounts to only few
optical cycles around 2.5 $\mu$m.

\section*{ACKNOWLEDGMENTS}
This work was supported by Austrian National Science Fund Project
M688.

  \end{document}